\documentclass[11pt,a4paper]{scrartcl}

\usepackage{CLICdp}
\usepackage{float}
\usepackage{amsmath}

\usepackage{CLICdp_definitions}

\RequirePackage{heppennames2}

\newcommand{\gHZZ}{\ensuremath{g_{\PH\PZ\PZ}}\xspace}
\newcommand{\gHWW}{\ensuremath{g_{\PH\PW\PW}}\xspace}

\title{The role of CLIC in Europe's course to the high-energy frontier}

\notitlestamp % Uncomment this line to remove the stamp with the CLICdp document number
\date{\today}
\addauthor{P.~N.~Burrows}{\institute{1}}
\addauthor{L.~Linssen}{\institute{2}}
\addauthor{A.~Robson}{\institute{3}}
\addauthor{D.~Schulte}{\institute{2}}
\addauthor{S.~Stapnes}{\institute{2}}
\addinstitute{1}{Unversity of Oxford, Oxford, United Kingdom}
\addinstitute{2}{CERN, Geneva, Switzerland}
\addinstitute{3}{University of Glasgow, Glasgow, United Kingdom}

\abstract{We respond to points raised in the recent discussion note arXiv:1912.13466, `Charting the European course to the high-energy frontier', which compares the CLIC and FCC programmes.}

\graphicspath{ {./logos/}{./figures/} }

\begin{document}

% generates the title page
\titlepage

% include source for sections
\section{CLIC and the European Strategy for Particle Physics}

The Compact Linear Collider (CLIC) \cite{clic-study} proposes \epem collisions up to multi-TeV energies, 
as documented in the Conceptual Design Report (2012) \cite{CLICCDR_vol1, CLIC_PhysDet_CDR} and recent comprehensive 
updates on the extensive accelerator, physics, and detector R\&D that has been carried out by the collaboration
\cite{staging_baseline_yellow_report, CLICsummaryYR, CLICPIPYR, deBlas:2018mhx, CLICDetectorTechYR}.
The CLIC baseline staging scenario includes operation at centre-of-mass energies of 380\,GeV, 1.5\,TeV, and 3\,TeV. 
The Future Circular Collider programme proposes several stages: \epem collisions up to 
a maximum of 365\,GeV (FCC-ee), with proton collisions at 100\,TeV in the longer term (FCC-hh) \cite{FCCCDRvol1,FCCCDRvol2,FCCCDRvol3}.

Following the European Strategy for Particle Physics Open Symposium in Granada,
a physics `Briefing Book' \cite{Strategy:2019vxc} has provided a 
performance comparison of various future collider options, and 
national representatives were invited to consider several options for the high-energy frontier
in Europe \cite{ESUBriefingBookSupportingNote}, 
including the complete CLIC programme, the complete FCC programme, and a combined scenario
in which \epem collisions are provided by CLIC, followed by proton collisions via FCC-hh.

In this note we respond briefly to several points raised in the recent discussion paper arXiv:1912.13466 \cite{Amaldi:2019yda},
`Charting the European course to the high-energy frontier', which compares the CLIC and FCC programmes.
First we address points made about the accelerator performance, followed by the physics sensitivity; and
finally we give a CLIC perspective on the European Strategy.

\section{Accelerator}

1912.13466 comments on the relative performance risk between the CLIC and FCC-ee accelerators.  In Granada it was concluded that with the current state of knowledge both appear feasible, based on past reviews.  Both push the performance well beyond the state of the art and explore new territory.  For example, in FCC-ee beamstrahlung will for the first time be important in a circular collider and can render the beam unstable, as has been discovered during the FCC study.  The CLIC beams at the collision point have a height of 3 nanometres at 380\,GeV; this is about ten times smaller than in FCC-ee, and the vertical beam quality, i.e. the normalized emittance, is about three times smaller.  In both colliders, these ambitious parameters are essential to achieve affordable cost and power consumption.  Experience with past and present accelerators and the technology developments give confidence that they can be met, as has been confirmed in detailed reviews of the projects.  1912.13466 mentioned the Stanford Linear Collider, SLC.  Actually, as the first of its kind SLC faced a number of limits that were analysed and overcome, e.g. by adding a large number of feedback systems.  The remaining limits have been understood and addressed in the CLIC design.  The most important SLC bunch charge limitation was due to a well understood instability in the damping ring.  A technology development and innovation programme has prototyped the required hardware to ensure nanometre-size beams for CLIC, and a test facility demonstrated the power production complex and beam acceleration.

Based on the important progress made since, both in rings and linacs, one should not be too concerned about either of the designs for \epem colliders at CERN. If we fail to be confident that by careful design, technical development, and implementation at CERN the performance of either of the two proposals can be pushed beyond the state of the art, then our field would have no future.

\section{Physics}

While also having a view to the longer term, the main focus of this European Strategy update is on choosing the next step.  Looking at the physics case for the next step:

\paragraph{Higgs physics:}
The Higgs physics case for the two machines FCC-ee and CLIC-380 is similar; more similar than presented in 1912.13466.  The comparison given there, of 8 years of data-taking at CLIC-380 and 14 years of data-taking at FCC-ee, is not reasonable.  A reasonable comparison is between FCC-ee and either the CLIC baseline of CLIC-380 + CLIC-1500 (15 years) or an extended run at CLIC-380 (e.g. 13 years) \cite{CLIClongerFirstStage} -- in either of these cases the Higgs performance is much more similar than that given in 1912.13466.  The cost of CLIC-380 is much less than that of FCC-ee365 (5.9\,BCHF versus 11.6\,BCHF).  The cost of CLIC-380 + CLIC-1500 (11.0\,BCHF) is similar to FCC-ee.

\paragraph{Top physics:}
The top-quark physics performance of CLIC-380 and FCC-ee365 is similar, but 1912.13466 does not mention that FCC-ee365 has twice the power consumption of CLIC380 (340 MW vs 168MW), as well as twice the cost.  Indeed 1912.11871 by Blondel \& Janot \cite{Blondel:2019ykp} notes that a circular machine is not ideally suited for reaching the top threshold.
Furthermore, at a linear collider, measurements with different beam polarisation, enriching the event samples in either left-handed or right-handed top quarks, give an additional handle for disentangling the photon and Z-boson contributions to the top-quark couplings.  The further option of going to a higher centre-of-mass energy at CLIC also has manifest benefits for top physics, giving a lever-arm to effectively constrain BSM EFT operators whose effects grow with energy \cite{Abramowicz:2018rjq}; indeed, a global interpretation of $\PQt\PAQt$ production using an EFT approach requires two energies.

\paragraph{Z/WW physics:}
A very high-statistics Z physics programme is indeed unique to FCC-ee, since a reconfigured CLIC can accumulate $5\times 10^9$ Z bosons compared with FCC-ee's $5\times 10^{12}$, though for the $A_f$ observables part of the statistical advantage is compensated at CLIC by electron beam polarisation.  One difference between FCC-ee and CLIC is the effect of the Z-pole programme on the extraction of Higgs couplings.  Without a Z-pole run, the precision on \gHZZ and \gHWW in the EFT fit of 1907.04311 \cite{deBlas:2019wgy} is similar for FCC-ee at 240 GeV and CLIC at 380 GeV. The Z-pole data provide further improvement at FCC-ee. In contrast, at CLIC the impact of the EW parameter uncertainties is largely mitigated by the higher-energy runs.

More generally, the option of going to higher centre-of-mass energies also gives CLIC much superior aTGC sensitivity compared with FCC-ee \cite{deBlas:2019rxi}, and opens interesting BSM reach, as has been extensively documented \cite{deBlas:2018mhx,Strategy:2019vxc}.

\subsection*{CLIC as the next step}

1912.13466 argues that the higher cost of FCC-ee compared with CLIC-380 is justified because part of the infrastructure for FCC-hh is included.  This neglects the point that CLIC-380 + FCC-hh has approximately the same cost (5.9+24=29.9\,BCHF) as FCC-all (28.6\,BCHF).  1912.13466 claims "Given the cost and effort needed for CLIC it will very likely preclude Europe from pursuing hadron collider physics beyond the LHC" -- this cost comparison shows that this is manifestly not the case.

Given the cost comparison, the question becomes whether the FCC-ee's Z/WW physics case \textit{alone} justifies embarking now on a programme that removes CERN's flexibility.

The option of CLIC-380 + FCC-hh must therefore not be ignored (as is done by 1912.13466).  Starting with CLIC-380 provides a much more flexible option -- preserving the possibility of moving to FCC-hh later, but also preserving the option of moving to higher CLIC energies; one extra motivation for that could be in the case where FCC-hh magnet technology remains in a state of unreadiness.  In addition, starting with CLIC-380 provides the possibility for concurrent progress in novel accelerator technologies for a potential future collider.

For the complete CLIC programme, 1912.13466 only reports the example that CLIC can discover Higgs (top) compositeness up to 10 (8) TeV, but neglects to mention that CLIC can reach new physics scales of 100\,TeV -- similar to FCC-hh but on a shorter timescale -- in some cases as reported in the Briefing Book \cite{Strategy:2019vxc}.

If alternatively FCC-hh is the long-term future, then all of the arguments given in 1912.13466 for needing FCC-ee Higgs and top results to fully exploit FCC-hh (top Yukawa, absolute \gHZZ determination etc) can equally be made for the sequence CLIC-380 + FCC-hh.

Furthermore, CLIC and FCC-hh are decoupled accelerator complexes, so can have a seamless handover; this is in contrast with FCC-ee, which has to be stopped to install FCC-hh.

1912.13466 argues that for CLIC, aiming at 10 to 100 times better accuracy will open a new range of systematic errors; the same point could equally be made about the FCC-ee programme, particularly with the statistics envisioned at the Z-pole.

\section{Strategy}

There seems to be widespread consensus that the next-generation collider should be an electron--positron collider.
Moving forward quickly with a linear collider (expandable, but with further stages not decided upon) 
would allow a vibrant high-energy frontier programme to be maintained over the coming decades, while 
pursuing in parallel the accelerator R\&D required to open future options.
This approach maintains maximal flexibility and keeps projects on `reasonable' timescales. 
The discussion of what will be the most appropriate high-energy frontier machine after
an initial \epem machine must be kept open such that it can be guided by new physics results, new technology,
and new generations of physicists and engineers.
An initial CLIC programme, in parallel with strong accelerator R\&D and HL-LHC, then followed by the
best possible high-energy frontier machine when technologies are mature, thus provides the most flexible
and appealing strategic option for collider physics in Europe.

% add references
\printbibliography[title=References]

\end{document}